\documentclass[prl,aps,twocolumn,showpacs,floatfix]{revtex4}
\usepackage{amsmath,amssymb,amsbsy}
\usepackage{graphicx}
\usepackage{dcolumn}
\usepackage{bm}
\let\oldphi\phi
\let\phi\varphi
\let\varphi\oldphi
\renewcommand{\deg}{{}^{\circ}}

\begin{document}

\title{Comment on ``Minimal size of a barchan dune''}

\author{B. Andreotti and P. Claudin}
\affiliation{
Laboratoire de Physique et M\'ecanique des Milieux H\'et\'erog\`enes,\\
UMR 7636 CNRS-ESPCI-P6-P7,\\
10 rue Vauquelin, 75231 Paris Cedex, France.}

\date{\today}

\begin{abstract}
It is now an accepted fact that the size at which dunes form from a flat sand bed as well as their `minimal size' scales on the flux saturation length. This length is by definition the relaxation length of the slowest mode toward equilibrium transport. The model presented by Parteli, Dur\'an and Herrmann [Phys. Rev. E \textbf{75}, 011301 (2007)] predicts that the saturation length decreases to zero as the inverse of the wind shear stress far from the threshold. We first show that their model is not self-consistent: even under large wind, the relaxation rate is limited by grain inertia and thus can not decrease to zero. A key argument presented by these authors comes from the discussion of the typical dune wavelength on Mars ($650$~m) on the basis of which they refute the scaling of the dune size with the drag length evidenced by Claudin and Andreotti [Earth Pla. Sci. Lett. \textbf{252}, 30 (2006)]. They instead propose that Martian dunes, composed of large grains ($500~\mu$m), were formed in the past under very strong winds. We show that this saltating grain size, estimated from thermal diffusion measurements, is not reliable. Moreover, the microscopic photographs taken by the rovers on Martian aeolian bedforms show a grain size of $87\pm25~\mu$m together with hematite spherules at millimetre scale. As those so-called ``blueberries'' can not be entrained by reasonable winds, we conclude that the saltating grains on Mars are the small ones, which gives a second strong argument against the model of Parteli~\emph{et al.}.
\end{abstract}


\maketitle
In the present comment, we adopt the point of view of Parteli, Dur\'an and Herrmann~\cite{PDH07} and use their model to point inconsistencies. We refer the interested reader to the series of papers published by the authors on the subject~\cite{ACD02,HDA02,A04,HEAACD04,ECA05,EAC07,CA06}.

\section{Modelling the saturation length}
The sand transport model used in \cite{PDH07} belongs to the series of models --~the `one species' models~-- in which one assumes that there is a single type of grain trajectories. The only self-consistent model of this type~\cite{A04} is that derived by Ungar and Haff \cite{UH87}, from which \cite{PDH07} is directly inspired. One assumes that the evolution of the sand flux is governed by the ejection process. Introducing the grain hop length $\ell$ and the number of ejected grains $N$ per unit impacting grain, one gets:
\begin{equation}
\ell \frac{dq}{dx} = N q.
\end{equation}
The fluid in the saltation curtain is assumed to be at equilibrium between the driving shear stress $\rho_f u_*^2$, the air-borne basal shear stress $\rho_f u_{\rm bas}^2$ and the sand-borne shear stress.  The sand-borne shear stress is proportional to the sand flux and to the difference between the velocity at which grains take off $v_\uparrow$ and collide back the sand bed $v_\downarrow$:
\begin{equation}
\rho_f u_*^2=\rho_f u_{\rm bas}^2+ \rho_s \frac{(v_\downarrow-v_\uparrow)}{\ell} q.
\end{equation}
At saturation, the wind is assumed to be just sufficient to maintain transport ($N=0$), which leads to a basal shear velocity $u_{\rm bas}$ independent of $u_*$ and thus equal to the threshold shear velocity $u_{\rm th}$. At saturation $\ell$, $v_\downarrow$ and $v_\uparrow$ are evaluated  in the saltation curtain, where the velocity profile is almost independent on $u_*$. The saturated flux can thus be put under the form:
\begin{equation}
q_{\rm sat}=\chi (u_*^2-u_{\rm th}^2).
\end{equation}
where $\chi$ depends only on the grain size, for a given atmosphere, but not on the wind strength.

Parteli \emph{et al.} then derive the saturation length by a simple linearization of the saturation equation under three assumptions. First, the number of ejected grains $N$ per unit impacting grain is assumed to be a function of the basal shear velocity $u_{\rm bas}$ only. Second, they assume that the grains ejected during collisions instantaneously reach the wind speed in the saltation curtain: the grains are assumed to have negligible inertia. Third, they assume that the wind instantaneously adjusts to changes of sand flux. One then gets:
\begin{equation}
\ell \frac{dq}{dx} = q_{\rm sat} \left. \frac{dN}{d u_{\rm bas}^2}\right|_{u_{\rm bas}=u_{\rm th}} \!\! \left(u_{\rm bas}^2-u_{\rm th}^2\right).
\end{equation}
This equation can be put under the form of a first order linear relaxation:
\begin{equation}
\ell_{\rm sat} \frac{dq}{dx} = q_{\rm sat}-q,
\end{equation}
where the saturation length is equal to:
\begin{equation}
\ell_{\rm sat} = \frac{\ell}{\frac{dN}{d u_{\rm bas}^2} (u_*^2-u_{\rm th}^2)} \, .
\end{equation}
Parteli \emph{et al.} have estimated the prefactor $\ell / \frac{dN}{d u_{\rm bas}^2}$  for a grain size of $250~\mu$m, on Earth, to $0.85$~m. As expected for any relaxation length, $\ell_{\rm sat}$ diverges at the threshold shear velocity. As $\ell$ does not depend on $u_*$, $\ell_{\rm sat}$ decreases as $1/u_*^2$ for large $u_*$.
\begin{figure}[t!]
\includegraphics{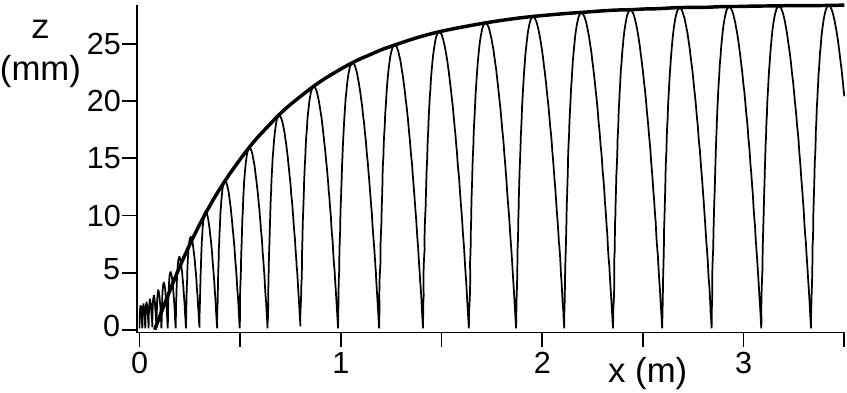}
\caption{Trajectory of a grain of $250~\mu m$ in the saltation curtain, on Earth. The transient length allows to define and measure the drag length $\ell_{\rm drag}$.}
\label{Trajectory}
\end{figure}

\begin{figure}[t!]
\includegraphics{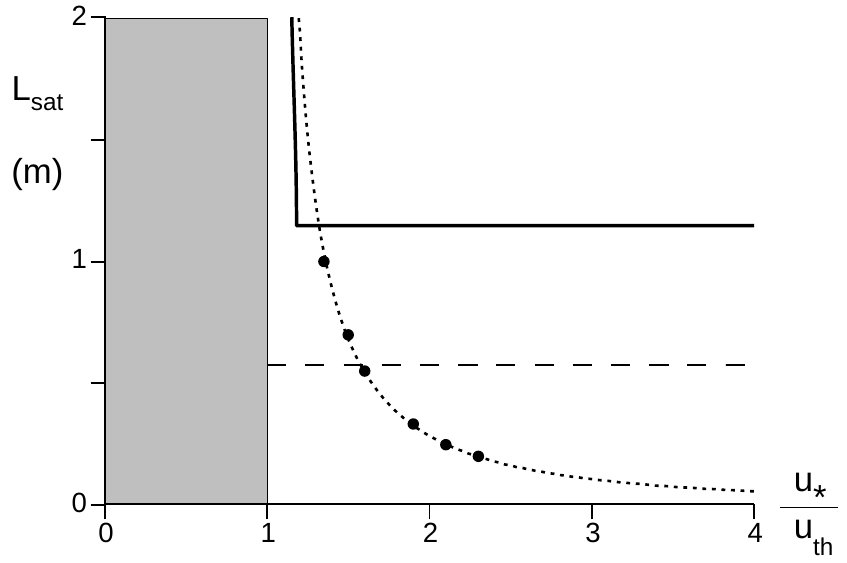}
\caption{Saturation length as a function of the rescaled wind shear velocity for sand grains of $250~\mu$m, on Earth. The dotted line corresponds to the model of Parteli \emph{et al.} (the symbols are deduced from fig.~5 of~\cite{PDH07}), that only takes into account the lag due to the ejection of grains. The solid line is the relaxation length obtained by modifying the model to take into account the grain inertia (Eq.~\ref{rouge}). The sharp transition is to be related to the second order dynamics. The dashed line shows the value of the drag length.}
\label{SaturationLength}
\end{figure}

In reality, there is not a single mechanism limiting the time and length of saturation transient but several~: (i) the ejection of grains, (ii) the grain inertia that controls the length needed for one grain to reach its asymptotic trajectory, (iii) the fluid inertia that controls the length needed for the wind to readapt to a change of $q$, (iv) the presence of grains above the saltation curtain with much longer trajectories \cite{A04}. It is worth emphasising that the ejection of grains (i) is the single source of lag considered in \cite{PDH07}. One should consider for $\ell_{\rm sat}$, the \emph{slowest} process i.e. the largest relaxation length amongst the modes of relaxation. We first wish to show that the saturation length proposed by Parteli \emph{et al.} is smaller than the relaxation length imposed by the grain inertia. The equation governing the grain motion may be written under the form:
\begin{equation}
\frac{d \vec v}{dt} =  \left(1-\frac{\rho_f}{\rho_s}\right) \vec g + \frac{3 \rho_f}{4 \rho_s d} \, C_d |\vec u-\vec v |~(\vec u-\vec v),
\end{equation}
where the drag coefficient is approximated by $$C_d=\left(\sqrt{C_\infty }+s \sqrt{\frac{\nu}{|\vec u-\vec v| d}} \right)^2$$. $C_\infty$ is the drag coefficient in the fully developed turbulent regime, i.e. at large particle Reynolds number.  In this limit, the drag length $\ell_{\rm drag}$, defined as the length needed for the grain to reach its asymptotic velocity, scales as $\rho_s/\rho_f \, d$. Consistently with Parteli \emph{et al.},  one can use for natural sand grains \cite{FC04,CA06} $C_\infty \simeq 1$ and $s \simeq 5$. Reasonable collisions rules are those considered in \cite{A04,CA06,PDH07}, with a restitution coefficient $e \simeq 0.6$ and a rebound angle around $45\deg$. Figure~\ref{Trajectory} presents the trajectory of an ejected grain submitted to a wind at $u_{\rm th}$, together with a fit of its envelop  by an exponential relaxation. The grain size is chosen to $d=250~\mu$m, for the seek of comparison with \cite{PDH07}. We find a drag length, of the order of $570$~mm. This means that the relaxation length associated to the ejection mechanism becomes smaller than that governed by grain inertia at moderately large velocity (around $u_*=1.5 u_{\rm th}$ in fig.~\ref{SaturationLength}).

The model of Parteli \emph{et al.} can be slightly modified to introduce the lag between the ejection of grains and the point at which they reach the saltation curtain velocity:
\begin{equation}
\ell \frac{dq}{dx} = \mathcal{Q} \quad{\rm and}\quad \ell_{\rm drag}\frac{d \mathcal{Q} }{dx} = N q-\mathcal{Q},
\label{rouge}
\end{equation}
where $\mathcal{Q}$ is the flux of grains just ejected and already accelerated by the wind. The saturation length, defined for this second order system as the slowest relaxation rate, then becomes:
\begin{equation}
\ell_{\rm sat} = \Re \left[\frac{2\ell_{\rm drag}}{1-\sqrt{1-4\frac{\ell_{\rm drag}}{\ell}~\frac{dN}{d u_{\rm bas}^2}~(u_*^2-u_{\rm th}^2)}}\right].
\end{equation}
It is plotted in figure~\ref{SaturationLength} together with the prediction by Parteli \emph{et al.}. One can see that the divergence of the saturation length at the threshold is due to the ejection process, as stated by Parteli \emph{et al.}. However, soon above the threshold (above $u_*=1.18 u_{\rm th}$ in fig.~\ref{SaturationLength}), there is at least another mechanism leading to a larger saturation length: the grain inertia.

We reach the first conclusion of this comment: as the saturation length cannot be smaller than the drag length, it cannot decrease with the wind strength far from the threshold. This is an evidence for the lack of self-consistence of the model proposed by Parteli \emph{et al.}. The grain inertia could well  be the limiting mechanism at large wind, as proposed in \cite{SKH01,ACD02,HDA02}, but this does not preclude the existence of even slower relaxation processes~\cite{A04,CA06}.
\begin{figure*}[t!]
\includegraphics{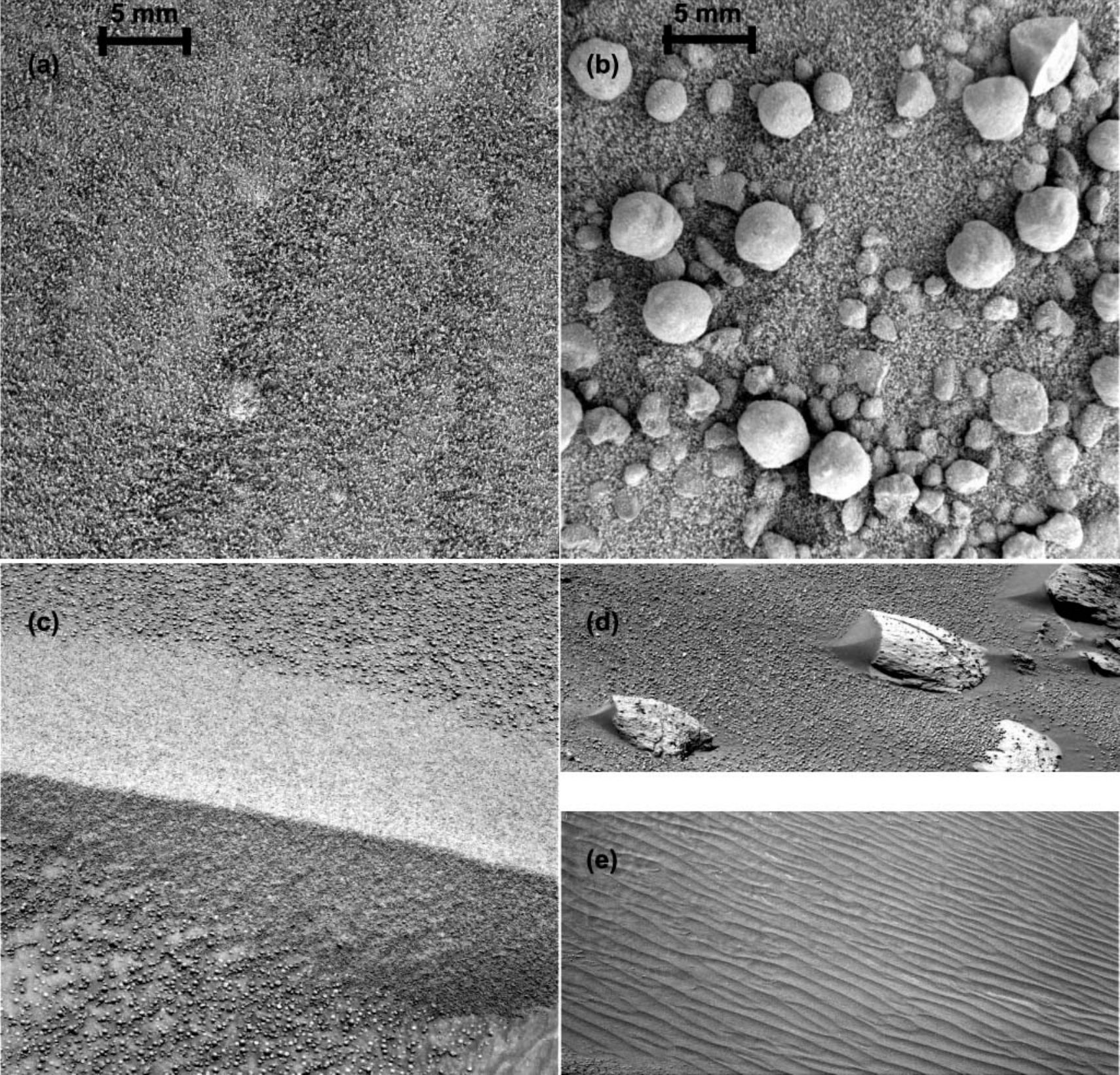}
\caption{{\bf a} Microscope photograph of the sand composing a Martian ripple. {\bf b} Microscope photograph showing the mixing of small grains and hematite spherules (``blueberries''), characteristic from the soil seen by the two rovers. {\bf c} Aeolian ripple on Mars, characteristic of transport in saltation. A strong difference of composition between the soil, covered by blueberries and the ripple can be observed. {\bf d} Aeolian nebkhas on Mars, characteristic of transport in saltation. These shadow dunes behind stones are clearly evidencing that small grains are transported in saltation, but not the hematite blueberries. {\bf e} Extended zone of aeolian ripples in a small scale impact crater. Blueberries may be seen at the bottom left of the picture, showing that the ripples are composed of small grains. These pictures have been taken by the rover Opportunity on Sols 58, 59, 60 and 85. They can be found on the NASA web site \texttt{http://marsrovers.jpl.nasa.gov/gallery/all/opportunity.html}. Courtesy NASA/JPL-Caltech.}
\label{MarsGrains}
\end{figure*}

\section{The size and density of grains on Mars}
The main argument presented by Parteli~\emph{et al.} in favour of a saturation length decreasing as the inverse of the wind shear stress comes from the typical size of Martian dunes. They use an estimated grain size $d\simeq 500\pm100~\mu$m derived by Edgett and Christensen~\cite{EC91} from the Viking Orbiter infrared thermal mapper (IRTM)  data. Using the simple scaling law of the dune wavelength based on $\ell_{\rm drag}$ \cite{CA06}, one would then expect with such a grain size a spacing of $4$~km between dunes on Mars. The real wavelength of Martian dunes is much smaller, between $500$~m and $700$~m \cite{CA06}. Parteli \emph{et al.} thus conclude that, in disagreement with our scaling relationship, very large winds are needed to explain the observed sizes i.e. to make the saturation length of large grains very small. We hereafter summarise the evidences given in \cite{CA06} that the grains in saltation on Mars are in fact much smaller than $500~\mu$m. 

As shown by Fenton \emph{et al.}~\cite{F03,FM06}, the determination of saltating grain size from thermal diffusion estimates is far to be obvious. Indeed, the measurement is very indirect. Edgett and Christensen~\cite{EC91} use the thermal model of Kieffer~\emph{et al.}~\cite{KMPJ77} to calculate thermal inertia with Viking IRTM data. Using the updated relation by Presley and Christensen~\cite{PC97}, the same data for the Hellespontus dunes give $1200\pm200~\mu$m instead of $500\pm100~\mu$m. If we push further the argument of Parteli \emph{et al.} with this new grain size, one would need a \emph{typical} shear velocity of $10$~m/s, which roughly corresponds to $500$~km/hour at $10$~m above the soil. This is more than one order of magnitude larger than present winds observed on Mars. In a dust devil rotating at such a speed, the depression in the core of the vortex would be larger than the average pressure of Martian atmosphere, which is not physically possible.

\begin{figure}[t!]
\includegraphics{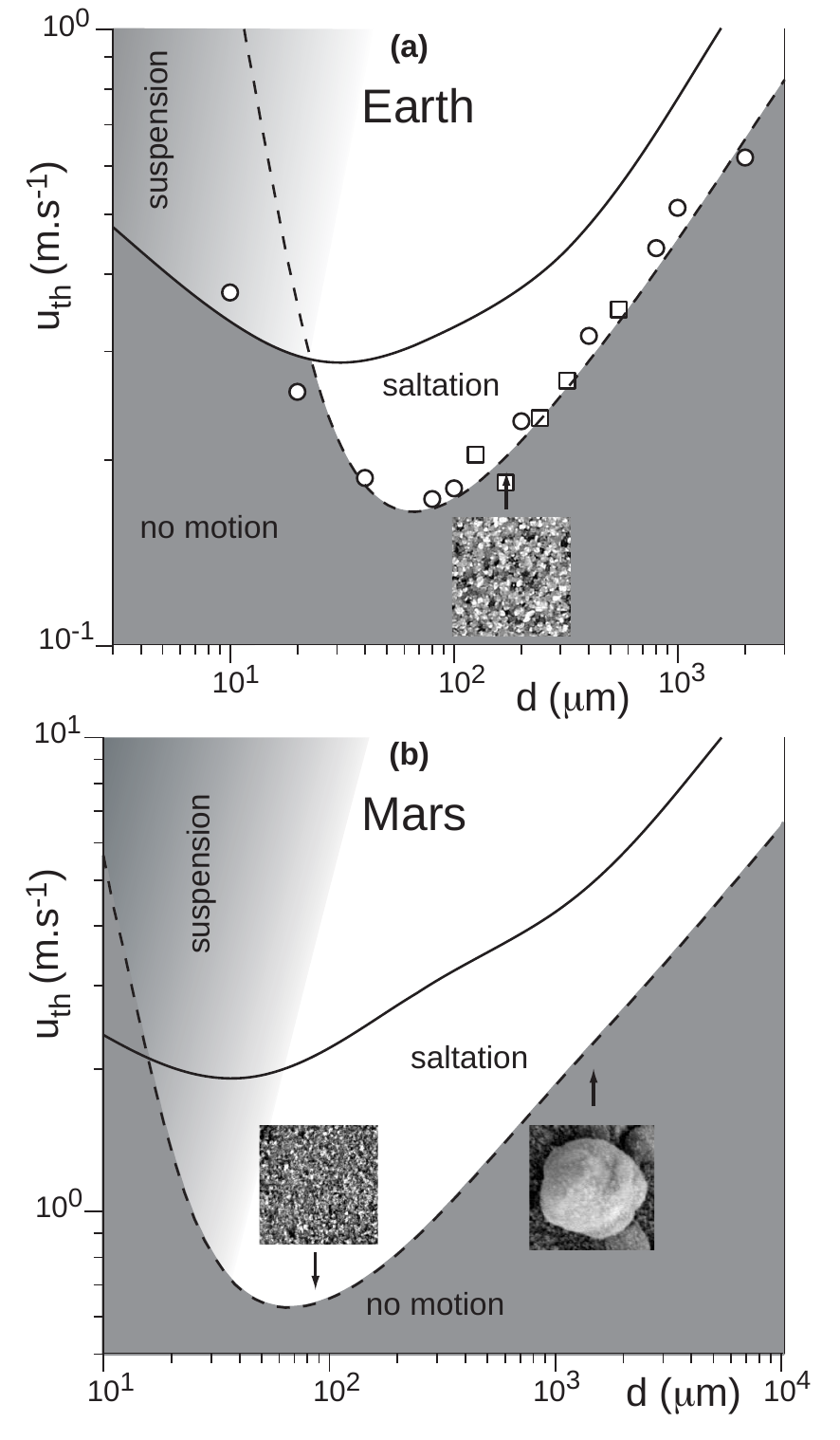}
\caption{Diagram showing the mode of transport on Earth {\bf (a)} and on Mars {(\bf b)}, as a function of the grain diameter $d$ and of the turbulent shear velocity $u_{\rm th}$. Below the dynamical threshold (dashed line), no grain motion is observed (dark gray). A grain at rest on the surface of the bed starts moving, dragged by the wind, when the velocity is above the static threshold (solid line). Between the dynamical and static thresholds, there is a zone of hysteresis where transport can sustain due to collision induced ejections. Above static threshold, the background color codes for the ratio $u_*/u_{\rm fall}$: white corresponds to negligible fluctuations and gray to suspension. The experimental points are taken from ({\Large$\circ$}) Chepil \cite{C45} and ($\square$) Rasmussen \cite{RIR96,A04} in the aeolian case. The insets shows the location of the observed grains in the diagrams.}
\label{ShieldsEarthMars}
\end{figure}

Much more reliable are the direct observations by the rovers Opportunity and Spirit. The photographs taken by the rovers (Figure~\ref{MarsGrains}a-b) mostly show two well separated grain sizes: large spherules of millimetric scale, composed of hematite ($\rho_s=5270$~kg/m$^3$) and small basalt grains with iron coating ($\rho_s=3010$~kg/m$^3$) between $60~\mu$m and $110~\mu$m \cite{CA06,JMGFW06}. Note that the thermal inertia measurements would point to large grains even in the zones where the rovers have found small ones~\cite{site}. How can one then discriminate between grains that can be transported in saltation and grains that cannot? The first argument is theoretical \cite{CA06}. With such large size and density, the threshold velocity for the entrainment of the blueberries into saltation is very large (fig.~\ref{ShieldsEarthMars}). The small grains, on the other hand, can be transported even with contemporary winds. It may be argued that the small grains would rather be transported into suspension but there is no clear threshold between saltation and suspension: as the wind speed increases, wind fluctuations become gradually more important with respect to gravity. The second argument comes from the field observations of aeolian structures on Earth. In particular, the formation of aeolian ripples and of shadow dunes behind obstacles (nebkhas) constitute a clear signature of transport into saltation. Figure~\ref{MarsGrains}c-e shows that these structures have a much higher concentration of small particles than the surrounding soil.

The emergent picture is thus very coherent: the grains transported in saltation on Mars are smaller than $100~\mu$m and are certainly not the millimetre scale hematite spherules blueberries; they can be transported by the present winds (ripples have formed in very recent impact craters; they form very recognisable aeolian bedforms like ripples and nebkhas, and probably dunes. The conclusions reached by Parteli \emph{et al.} are thus probably wrong, for the particular problem of dune formation on Mars and for the modelling of sand flux saturation transients in general.

\section{Relation between the wavelength at which dunes form and the saturation length}
The instability of a flat sand bed results from the interaction between the sand bed profile, which modifies the fluid velocity field, and the flow that modifies in turn the sand bed as it transport grains. The fluid is accelerated on the upwind (stoss) side of proto-dunes and decelerated on the downwind side. This results into an increase of the shear velocity $u_*$ applied by the flow on the stoss side of the bump. Conversely, $u_*$ decreases on the lee side. As the saturated sand flux is an increasing function of $u_*$, erosion takes place on the stoss slope as the flux increases, and sand is deposited on the lee of the bump. If the velocity field was symmetric around the bump, the transition between erosion and deposition would be exactly at the crest, and this would lead to a pure propagation of the bump, without any change in amplitude (the `$A$' effect). In fact, due to the simultaneous effects of inertia and dissipation, the velocity field is asymmetric (even on a symmetrical bump) and the position of the maximum shear stress is shifted upwind the crest of the bump (the `$B$' effect). In addition,  the sand transport reaches its saturated value with a spatial lag $\ell_{\rm sat}$. The maximum of the sand flux $q$ is thus shifted downwind  the point at which $u_*$ is maximum by a typical distance of the order of $\ell_{\rm sat}$. The criterion of instability is then geometrically related to the position at which the flux is maximum with respect to the top of the bump: an up-shifted position leads to a deposition of grains before the crest, so that the bump grows.

These arguments can be formalized by performing the linear stability analysis of a flat sand bed  \cite{ACD02,ECA05}. For a small deformation of the bed profile $h(t,x)$, the excess of stress induced by a non-flat profile can be written in Fourier space as $\rho_f u_*^2 (A + iB) k \hat{h}$. $A$ and $B$ may be in principle deduced from a turbulent closure. Jackson and Hunt~\cite{JH75,KSH02} have derived asymptotic expressions for $A$ and $B$ as functions of $\ln(kz_0)$, where $z_0$ is the aerodynamic roughness.
\begin{figure}[t!]
\includegraphics{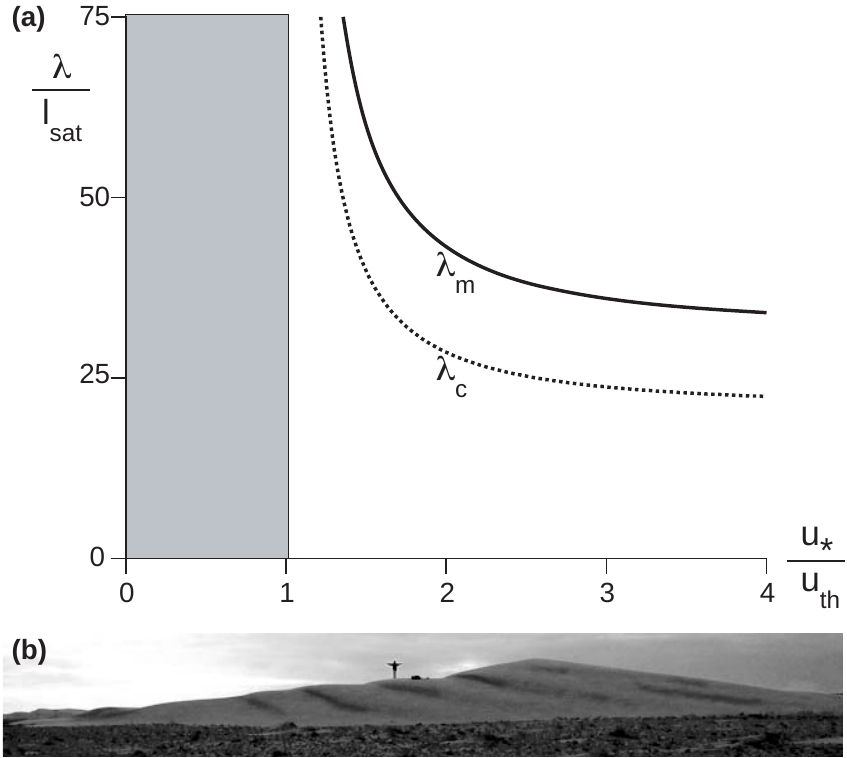}
\caption{{\bf a} Relation between the wavelength $\lambda_m$ at which dunes appear and the saturation wavelength, as a function of the wind strength.  The limit of stability $\lambda_c$ (wavelength for which $\sigma=0$) is also shown. The plot has been produced for $A=5$ and $B=1.5$, which are typical values predicted by Jackson and Hunt \cite{JH75,KSH02}. The slope effect, due to gravity, has an increasing importance close to the threshold. {\bf b} Destabilization of the flanks of a barchan dune during a violent dust storm coming from Sahara towards Canarias, in April 2003. The wavelength of destabilization is reduced by a rough factor of $2$ with respect to that observed during regular trade winds ($\simeq20~m$).}
\label{Instability}
\end{figure}

At this stage, for who wishes to catch subdominant dependencies on the wind shear velocity, there is again a very important mechanism forgotten in Parteli \emph{et al.}: the influence of the slope on the threshold shear stress. As shown by Rasmussen \emph{et al.}~\cite{RIR96}, at the linear order, the threshold shear stress may be written as: $\rho_f u_{\rm th}^2 (1+\partial_x h/\tan \theta_a)$, where $\theta_a \simeq 32\deg$ is the avalanche repose angle. This gravity effect originates from the trapping of grains at the surface of the sand bed, influenced by the slope. In the Fourier space, the saturated flux modulation can be written as:
\begin{equation}
\hat q_{\rm sat}=\left[ (A + iB)u_*^2  -i u_{\rm th}^2/\tan \theta_a \right] \chi  k\hat{h}.
\end{equation}
Using a first order linear saturation equation and the conservation of matter, we end with a growth rate $\sigma$ of the form:
\begin{equation}
\sigma=\frac{\chi u_*^2 k^2}{1+k^2l_{\rm sat}^2}~\left[ B  - \frac{u_{\rm th}^2}{ u_*^2~\tan \theta_a} -A k l_{\rm sat}\right].
\end{equation}
Figure~\ref{Instability} shows the relation between the wavelength $\lambda_m$ at maximum growth rate and the marginally stable wavelength $\lambda_c$, as functions of the rescaled wind shear velocity $u_*/u_{\rm th}$. It can be observed that these wavelengths decrease with wind strength, due to the decreasing relative importance of gravity effects with respect to wind effects. This slope effect could be in fact the dominant explanation for the observed variations of minimal size with the wind strength (typically a factor of $2$ in Morocco, see figure~\ref{Instability}).

In conclusion, if the role of the particle diameter and of the fluid to grain density ratio on the time and length scales of dunes  is now pretty clear, that of wind speed remains controversial. Further work is needed to shed light on the influence of the numerous dynamical mechanisms involved.

\end{document}